\documentclass{article}

\begin{document}

\title{Capillarity driven spreading of circular drops of shear-thinning 
fluid}

\author{S.I. Betel\'{u}\\ 
Mathematics Department\\
University of North Texas\\
P.O. Box 311430, Denton, TX 76203-1430\\
\\M.A. Fontelos\\
Departamento de Matem\'{a}tica Aplicada\\
Universidad Rey Juan Carlos\\
M\'ostoles, Madrid 28933}

\date{ May 17 2002}

\maketitle

\begin{abstract}
We investigate the spreading of thin, circular liquid drops of power-law
rheology. We derive the equation of motion using the thin film
approximation, construct source-type similarity solutions and compute the
spreading rate, aparent contact angles and height profiles. In 
contrast with the spreading of newtonian liquids, the contact line paradox does not arise for
shear thinning fluids.
\end{abstract}

In this work we study the spreading of circular drops of power-law rheology fluids, 
also known as Ostwald-de Waele fluids \cite{Bird}. The
power-law rheology is one of the simplest generalizations of the Newtonian
one, in which the effective viscosity is assumed to be a power law of 
the local rate of deformation $\dot \gamma$ as $\mu= m |\dot \gamma|^{1/\lambda
-1}$.
The values of $m$ and $\lambda$ depend on the physical
properties of the liquid. When $\lambda>1$, the fluid is called {\it shear-thinning}
and the viscosity tends to zero at high strain rates \cite{Bird}.

The problem of drop spreading has been intensively studied in the last
decades (see \cite{Myers,Greenspan,Bernis,Bertozzi,GrattonR}). The
motivation is that this class of flows plays a very important
role in processes such as coating and painting. One reason for our study is that 
while the past work considers only newtonian fluids, most of the
fluids of technological interest are non-newtonian. 

There is also a theoretical motivation:
one modelling difficulty for newtonian fluids is {\it the paradox of the contact line},
that states that the dissipation of energy is unbounded near the advancing contact line of 
a newtonian fluid.
This paradox may be solved, for example, by introducing the effect of the intermolecular forces, 
or abandoning the no-slip boundary condition at the substrate.
Here we show the remarkable fact that for power-law fluids with $\lambda>1$ the 
paradox does not arise. Moreover, there are solutions with moving interfaces, and their energy
dissipation rate is bounded. The qualitative reason lies in the fact that
the flow creates a layer adjacent to the substrate where the strain rate
diverges, and thus, for $\lambda >1$ the effective viscosity $\eta$ tends to
zero. In other words, the drop  self-lubricates.

Spreadings of power law fluids were first studied experimentally in \cite{Carre}
and the one dimensional flow was analyzed in \cite{AML,King}.
Here we construct for the first time radially symmetric spreading solutions, which can
be used directly to model droplet spreading. 

In power-law fluids, the deviatoric stress tensor $\tau _{ij}$ is related to the strain
tensor according to the following constitutive relation \cite{Bird,Gratton}: 
\begin{equation}
\tau _{ij}=m|\dot{\gamma}|^{1/\lambda -1}\dot{\gamma}_{ij},
\label{constitutive}
\end{equation}
where $|\dot{\gamma}|=\sqrt{\frac 12\sum_{i,j}\dot{\gamma}_{ij}\dot{\gamma}_{ij}}$ 
and $\dot{\gamma}_{ij}$ is the strain tensor, given by 
\begin{equation}
\dot{\gamma}_{ij}=\frac{\partial v_i}{\partial x_j}+\frac{\partial v_j}{%
\partial x_i}. 
\end{equation}
Here $v_i$ is the fluid velocity field. When $\lambda =1$, one recovers a
Newtonian fluid. If $\lambda >1$, the fluid is said to be {\it shear-thinning%
}, which is the most common case that includes many polymer solutions. The
values of $\lambda$ are typically between $1.7$ and $6.7$ \cite{Bird}.

In order to derive the equation of motion we use the {\it lubrication approximation}, 
which is widely applied in the case of Newtonian flows (for which $\lambda =1)$ (see for
instance \cite{GrattonR,Myers} and references therein).
We assume that the film is much thinner than its horizontal dimension, that the
motion is nearly horizontal and that the inertial effects are negligible so
that the flow is governed by a balance between capillary and viscous forces.
We use a cylindrical coordinate system $(r,\theta ,z)$ and assume 
circular symmetry. In this system, 
the velocity of the fluid has components $(u(r,z,t),0,v(r,z,t))$ and the substrate 
is at $z=0$. 
We neglect systematically the $z$ component of the velocity when it is 
compared with the $r$ component, $|u|\gg |v|$. 
We also assume that the stresses are mainly due to
high gradients of the horizontal velocity $u$ in the $z$ direction.
Consistently, we assume that the components $\dot{\gamma}_{rz}$ and 
$\tau _{rz}$ are much larger than all its other respective components. 
Then, the $r$-component of the momentum equation is reduced to 
\begin{equation}
-\frac{\partial p}{\partial r}+\frac{\partial \tau _{rz}}{\partial z}=0
\label{forcer}
\end{equation}
where $p$ is the pressure. Let $z=h(r,t)$ be the fluid free surface. The
conservation of mass can be written within the lubrication approximation as
\begin{equation}
h_t+\frac 1r(rUh)_r=0,  \label{masse}
\end{equation}
where $U$ is the horizontal velocity averaged in the $z$ coordinate. 
We assume that the drop size is much smaller than the capillary length. 
Then we can neglect gravity and 
the pressure under the free surface may be approximated by 
\begin{equation}
p(r,t)=\gamma \kappa +p_0 \label{pressure}
\end{equation}
where $\gamma$ is the surface tension, $\kappa$ is the mean curvature of the free surface
and $p_0$ the atmospheric pressure.
Under the lubrication approximation the curvature can be approximated by
\begin{equation}
    \kappa= -h_{rr}-\frac{h_r}{r}   \label{k}.
\end{equation}
One can integrate Eq. \ref{forcer} with respect to $z$ to obtain the shear
stress 
\begin{equation}
\tau _{rz}=p_r(z-h),  \label{stressint}
\end{equation}
that satisfies the zero-tangential stress condition at the free surface $z=h(r,t)$.
By using the constitutive
relation given by Eq. \ref{constitutive}, 
\begin{equation}
\tau _{rz}=m|u_z|^{1/\lambda -1}u_z, 
\end{equation}
and by substituting $\tau _{rz}$ from Eq. \ref{stressint} we can compute the 
$z$ derivative of the horizontal velocity, 
\begin{equation}
u_z=-m^{-\lambda }|p_r|^{\lambda -1}p_r(h-z)^\lambda . 
\end{equation}
After integration and using the no-slip condition\\ $u(z=0)=0$ one obtains 
\begin{equation}
u=m^{-\lambda }|p_r|^{\lambda -1}p_r\left( \frac{(h-z)^{\lambda +1}}{\lambda
+1}-\frac{h^{\lambda +1}}{\lambda +1}\right) . 
\end{equation}
Now, in order to apply the equation for the conservation of the mass Eq. \ref
{masse}, we need the averaged velocity $U$, 
\begin{equation}
U=\frac 1h\int_0^h{u}dz=-m^{-\lambda }|p_r|^{\lambda -1}p_r\frac{h^{\lambda
+1}}{\lambda +2}.  \label{ave}
\end{equation}
Finally, using Eq. \ref{masse} we obtain the equation of motion 
\begin{equation}
h_t-\frac 1r\frac 1{\lambda +2}\left( \frac \gamma m\right) ^\lambda
(rh^{\lambda+2}|\kappa _r|^{\lambda -1}\kappa _r)_r=0,  \label{sistema}
\end{equation}
which forms a system of partial differential equations with Eq \ref{k}.

Now we solve the problem in which a finite volume $V$ of fluid is initially
concentrated on a point in a planar surface. As there are no external length-scales
in this problem, we shall look for similarity solutions of the first kind 
\cite{Barenblatt} 
\begin{equation}
h(r,t)=\frac B{t^{2\beta }}H\left( \frac r{At^\beta }\right)  \label{ans}
\end{equation}
where $A$ and $B$ are constants and $\beta $ is the {\it similarity exponent}.
The argument of $H$ is the similarity variable $\eta =r/At^\beta$. If the
function $H(\eta)$ has a zero at $\eta _f$
then the front of the spreading will be given by 
\begin{equation}
r_f(t)=A\eta _ft^\beta . \label{rf0}
\end{equation}
Eq. \ref{ans} guarantees that the fluid volume $V$ is conserved, i.e. that 
\[V=\int_0^{r_f(t)}2\pi rh(r,t)dr=A^2B I
=const.\]
where we define the {\it shape factor}, 
\[
I= \int_0^{\eta _f}2\pi \eta H(\eta )d\eta.
\]
It is convenient to introduce the dimentionless curvature
\begin{equation}
K= -H'' - \frac{H'}{\eta} \label{Kl}
\end{equation}
such that, from Eq. \ref{k}, $\kappa (r,t)=K(\eta) V/(A^4 I t^{4\beta})$.
By substituting Eq. \ref{ans} in Eq. \ref{sistema} we find that the
similarity exponent is given by 
\[
\beta =\frac 1{7\lambda +3}. 
\]
By writing $A^{7\lambda +3}= \frac{7\lambda+3}{\lambda+2} (\gamma/ m)^\lambda
(V/I)^{2\lambda+1}$
and integrating once, we obtain the ODE
\begin{equation}
-\eta H=  H^{\lambda+2} |K'|^{\lambda-1} K' 
\label{selfsimilar1}
\end{equation}
which has to be integrated with Eq. \ref{Kl}.
It is clear that for $\eta \ge 0$ and $H \ge 0$, $K' \le 0$ and then
the equation can be written as
\begin{equation}
\eta H=  H^{\lambda+2} (-K')^{\lambda}. 
\label{selfsimilar}
\end{equation} 
The initial conditions are
\begin{equation}
H(0)=1, \hspace{1.cm} H'(0)=0, \hspace{1.cm} K(0)=\kappa_0. \label{bc}
\end{equation}
The first condition represents the scaled height at the center of the drop,
the second is the radial symmetry condition, and the third is the central
value of the curvature, which is a free parameter $\kappa_0$ 
that has to be calculated.

Using Eq. \ref{rf0} we obtain
a simple formula to compute the radius of the drop as a function of
the volume $V$, $\gamma$ and $m$:
\begin{equation}
r_f(t)= S_\lambda \left( V^{2\lambda+1} (\gamma/ m)^\lambda t\right)^{1/(7\lambda+3)} .
\label{rf}
\end{equation}
where $S_\lambda= ((7\lambda+3)/(\lambda+2))^{1/(7\lambda+3)} \eta_f /I^{(2\lambda+1)/(7\lambda+3)}$ 
depends only on $\lambda$.

The aparent contact angle of the drop is defined as $\tan \theta(t)=-h_r(r_i,t)$, where 
$r_i$ is the point of maximum slope, $h_{rr}(r_i,t)=0$. By simple substitution,
\begin{equation}
\tan \theta(t)= Q_\lambda
V^{\lambda/(7\lambda+3)}(m/\gamma)^{3\lambda/(7\lambda+3)}t^{-3/(7\lambda+3)}
\label{theta}
\end{equation}
where $Q_\lambda=-H'(\eta_i)
((\lambda+2)/(7\lambda+3))^{3/(7\lambda+3)}I^{-\lambda/(7\lambda+3)}$
and $\eta_i$ is the point where $H''(\eta_i)=0$.

We solve Eqs. \ref{Kl} and \ref{selfsimilar} subject to Eq. \ref{bc} 
numerically with a fourth order Runge-Kutta
scheme, with a variable stepsize equal to $H/10000$. 
In Fig. 1 we show several height profiles for different values of $\kappa_0$
with $\lambda=5/2$ fixed. For $\kappa_0<3.00428$ the solution is 
always positive: this solution does not represent a drop. 
For $\kappa_0>3.00428$, the drop profile tends to zero linearly at $\eta=\eta_f$. 
For a critical value of $\kappa_0 \approx 3.00428$ the profile tends to zero with
zero contact angle. This case represents a drop of a liquid that wets the
surface.
A similar picture is obtained for all $\lambda>1$. In general, there is a solution with zero
contact angle for a critical value of $\kappa_0=\kappa_\lambda$ and a continuum of other
solutions with
finite
contact angle for $\kappa_0>\kappa_\lambda$. In this sense, $\kappa_\lambda$ is an eigenvalue for
zero contact angle drops.

From the numerical calculations we observe that
the values of $S_\lambda$ are maxima for the solutions with zero contact angle.
That means that for fixed $V$, $m$ and $\gamma$, the zero contact angle drops have maximum
spreading
speed (see Eq. \ref{rf}).

In Fig. 2 we show the solution with zero contact angle for several values of 
$\lambda= 1.1, 1.5, 2, 3, 4$ and $5$. The region where the slope tends to zero near the front
is very small. 
There, the asymptotic expansion of the solution
with zero contact angle is $H(\eta)=$
\begin{eqnarray}
\left(\frac{(2\lambda+1)^3}{3\lambda(\lambda-1)(\lambda+2)}\right)^
{\lambda/(2\lambda+1)}\eta_f^{1/(2\lambda+1)}
(\eta_f -\eta)^{3\lambda/(2\lambda+1)} \label{asymp} \\
+ \mbox{lower order terms}. \nonumber
\end{eqnarray}

In table 1 we give the values of $S_\lambda$ and $Q_\lambda$ as a function of $\lambda$. They may be useful for future experimental
comparisons using Eq. \ref{rf} and Eq. \ref{theta}.
\begin{table}\hspace{1.5cm}
\begin{tabular}{cccccc}
$\lambda$   &   $\kappa_\lambda$& $\eta_f$ &  I       & $S_\lambda$ & $Q_\lambda$\\
1.1         &   7.64300         & 0.73220  &  0.83124 & 0.86880     & 1.814294 \\
1.5         &   4.08651         & 1.03906  &  1.61381 & 0.99651     & 1.137746 \\
2.0         &   3.29562         & 1.19915  &  2.08030 & 1.05262     & 0.951946 \\
2.5         &   3.00428         & 1.28882  &  2.34984 & 1.08074     & 0.878580 \\
3.0         &   2.85609         & 1.34757  &  2.52782 & 1.09766     & 0.840258 \\
4.0         &   2.71025         & 1.42076  &  2.74976 & 1.11683     & 0.802190 \\
5.0         &   2.64043         & 1.46484  &  2.88307 & 1.12721     & 0.784121 \\
\end{tabular}
\caption{Constants defining the selfsimilar solutions.}
\end{table} 

Finally, within the lubrication approximation, the dissipation rate of energy is
\begin{equation}
D=  \frac{2\pi}{\lambda+2} \gamma \left(\frac{\gamma}{m}\right)^\lambda   \int_0^{r_f} r |k_r|^{\lambda+1}
 h^{\lambda+2} dr  
\end{equation}
and for our self similar solution, it is proportional to 
\begin{equation}
2\pi \int_0^{\eta_f} \eta |K'|^{\lambda+1} H^{\lambda+2} d\eta.
\end{equation}
and then, using Eq. \ref{asymp} it is easy to see that the local rate of dissipation of energy 
converges near the front. As a consequence, in contrast with a newtonian fluid, the paradox of the 
contact line does nor arise for power law fluids with $\lambda>1$.

\newpage
LIST OF FIGURES\\

Fig 1- Solutions for $\lambda=5/2$ and $\kappa=4, 3.00428$ and $2$ (from left to right).\\

Fig 2- Zero contact angle solutions for $\lambda=1.1, 1.5, 2.0, 3.0, 4.0$ and $5.0$ (from left to
right).

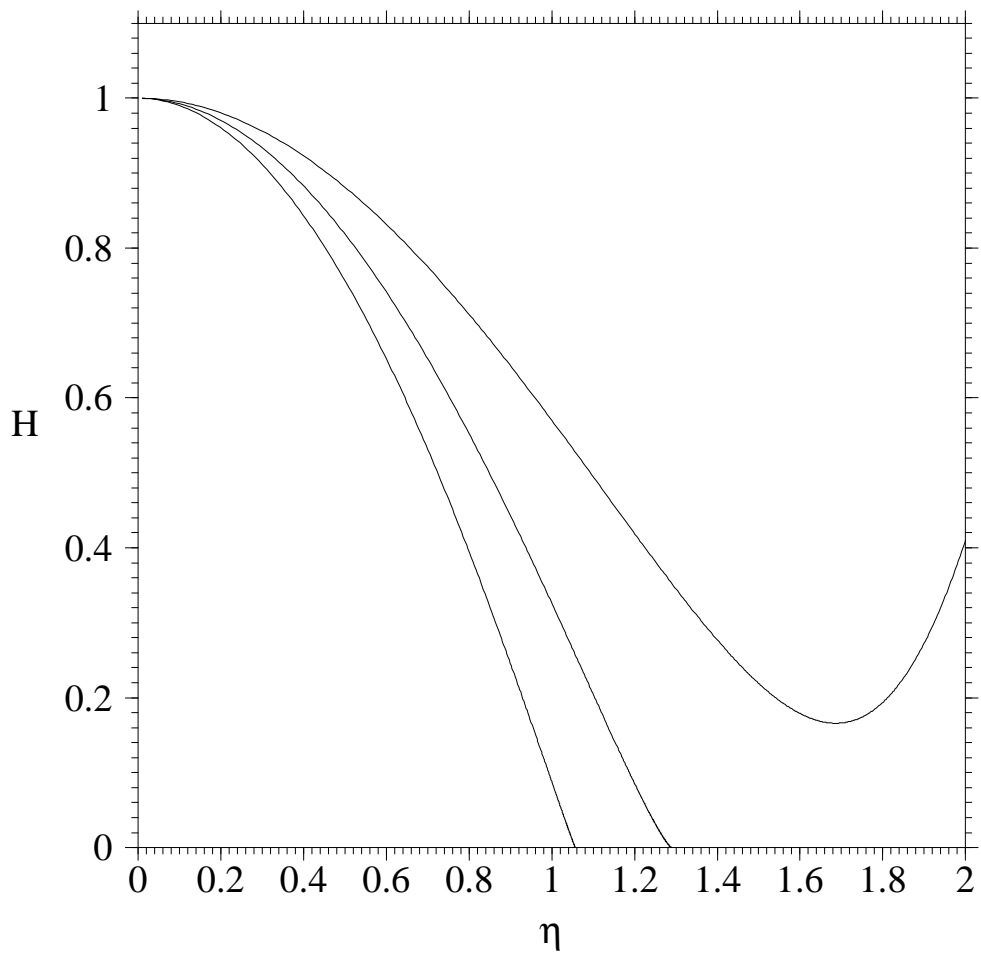
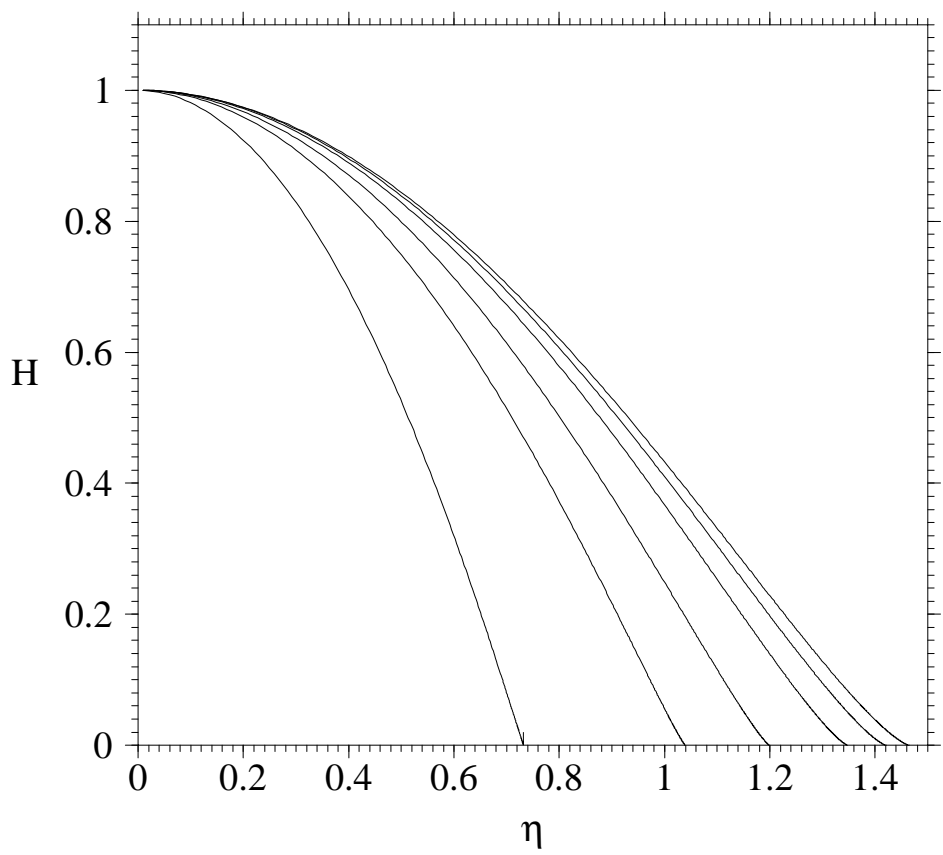
\end{document}